\documentclass[prd,onecolumn,nofootinbib,showkeys]{revtex4}
\newcommand\gothfamily{\usefont{U}{ygoth}{m}{n}}
\DeclareTextFontCommand{\textgoth}{\gothfamily}

\begin{document}

\title{COSMOLOGY WITH TORSION: AN ALTERNATIVE TO COSMIC INFLATION}

\author{{\bf Nikodem J. Pop{\l}awski}}

\affiliation{Department of Physics, Indiana University, Swain Hall West, 727 East Third Street, Bloomington, Indiana 47405, USA}
\email{nipoplaw@indiana.edu}

\noindent
{\em Physics Letters B}\\
Vol. {\bf 694}, No. 3 (2010) pp. 181--185\\
\copyright\,Elsevier B. V.
\vspace{0.4in}

\begin{abstract}
We propose a simple scenario which explains why our Universe appears spatially flat, homogeneous and isotropic.
We use the Einstein-Cartan-Kibble-Sciama (ECKS) theory of gravity which naturally extends general relativity to include the spin of matter.
The torsion of spacetime generates gravitational repulsion in the early Universe filled with quarks and leptons, preventing the cosmological singularity: the Universe expands from a state of minimum but finite radius.
We show that the dynamics of the closed Universe immediately after this state naturally solves the flatness and horizon problems in cosmology because of an extremely small and negative torsion density parameter, $\Omega_S\approx-10^{-69}$.
Thus the ECKS gravity provides a compelling alternative to speculative mechanisms of standard cosmic inflation.
This scenario also suggests that the contraction of our Universe preceding the bounce at the minimum radius may correspond to the dynamics of matter inside a collapsing black hole existing in another universe, which could explain the origin of the Big Bang.
\end{abstract}


\maketitle

According to the current model of cosmology, the Universe, shortly after the Big Bang, has undergone a very brief period of an extremely rapid, exponential acceleration known as cosmic inflation \cite{inf1,inf2}.
Cosmic inflation explains why the Universe appears spatially flat and why it is homogeneous and isotropic on large scales \cite{inf1}.
However, inflationary scenarios require the existence of scalar fields or other speculative and not-fully-understood mechanisms which introduce additional parameters to this model \cite{inf2}.
Moreover, Big Bang cosmology does not explain the origin of the initial, extremely hot and dense state of the Universe and what existed before this state.

Here we show that extending Einstein's general relativity to include the intrinsic angular momentum (spin) of matter, which leads to the Einstein-Cartan-Kibble-Sciama (ECKS) theory of gravity \cite{Kib,Sci1,Sci2,HD,HHKN,Lord,Sha,Ham,Niko0}, naturally explains why the Universe is spatially flat, homogeneous and isotropic, without invoking inflation.
We also propose that the torsion of spacetime, which is produced by the spin of quarks and leptons filling the Universe and prevents the formation of singularities (points of spacetime with infinite curvature and matter density), provides a physical mechanism for a scenario in which each collapsing black hole gives birth to a new universe inside it.
Such an attractive scenario of fecund universes has been proposed earlier by Smolin, however, either assuming the formation of singularities and linking the final singularity in each black hole to the initial singularity of a new universe, or avoiding a singularity but without explaining such an avoidance \cite{Smo1,Smo2}.
Torsion thus appears as a very plausible phenomenon: it provides a theoretical description of how a collapsing black hole produces a new nonsingular universe in the above scenario, explaining the origin of our Universe as the interior of a black hole existing in another, bigger universe \cite{Pat,Stu} and the arrow of time, and eliminates the need for inflation in cosmology.

\section{ECKS gravity}
The ECKS theory of gravity naturally extends Einstein's general relativity to include matter with spin, providing a more complete account of local gauge invariance with respect to the Poincar\'{e} group \cite{Kib,Sci1,Sci2,HD,HHKN,Lord,Sha,Ham,Niko0}.
It is a viable theory, which differs significantly from general relativity only at densities of matter much larger than the nuclear density.
This theory is advantageous over general relativity because torsion appears to prevent the formation of singularities from matter composed of particles with half-integer spin and averaged as a spin fluid \cite{Kop,Tra,HHK,dSS1,dSS2}, and to introduce an effective ultraviolet cutoff in quantum field theory for fermions \cite{Niko2}.

The ECKS gravity is based on the Lagrangian density of the gravitational field that is proportional to the Ricci curvature scalar $R$, as in general relativity.
However, this theory removes the general-relativistic restriction of the symmetry of the affine connection $\Gamma^{\,\,k}_{i\,j}$, that is, of the vanishing of the torsion tensor $S^k_{\phantom{k}ij}=(\Gamma^{\,\,k}_{i\,j}-\Gamma^{\,\,k}_{j\,i})/2$.
Instead, the torsion tensor is regarded as a dynamical variable, in addition to the metric tensor $g_{ij}$ \cite{Kib,Sci1,Sci2,HHKN,Lord}.
Varying the total action for the gravitational field and matter with respect to the metric gives the Einstein field equations that relate the curvature of spacetime to the canonical energy-momentum tensor of matter $\sigma_i^{\phantom{i}j}=\Theta_i^{\phantom{i}j}/\sqrt{-\mbox{det}\,g_{mn}}$ (we use the notation of \cite{Niko2}):
\begin{equation}
R_{ik}-\frac{1}{2}Rg_{ik}=\kappa\sigma_{ki},
\label{Ein}
\end{equation}
where $R_{ik}$ is the Ricci tensor of the Riemann-Cartan connection
\begin{equation}
\Gamma^{\,\,k}_{i\,j}=\{^{\,\,k}_{i\,j}\}+S^k_{\phantom{k}ij}+S_{ij}^{\phantom{ij}k}+S_{ji}^{\phantom{ji}k},
\label{con}
\end{equation}
$\{^{\,\,k}_{i\,j}\}$ are the Christoffel symbols of the metric $g_{ij}$, and $\kappa=8\pi G/c^4$.
Varying the total action with respect to the torsion gives the Cartan field equations that relate algebraically the torsion of spacetime to the canonical spin tensor of matter $s_{ij}^{\phantom{ij}k}=\Sigma_{ij}^{\phantom{ij}k}/\sqrt{-\mbox{det}\,g_{mn}}$:
\begin{equation}
S_{jik}-S^l_{\phantom{l}il}g_{jk}+S^l_{\phantom{l}kl}g_{ji}=-\frac{1}{2}\kappa s_{ikj}.
\label{Car}
\end{equation}

The symmetric dynamical energy-momentum tensor of general relativity $T_{ik}$ \cite{LL} is related to the canonical energy-momentum tensor by
\begin{equation}
T_{ik}=\sigma_{ik}-\frac{1}{2}(\nabla_j-2S^l_{\phantom{l}jl})(s_{ik}^{\phantom{ik}j}-s_{k\phantom{j}i}^{\phantom{k}j}+s^j_{\phantom{j}ik}),
\label{sym}
\end{equation}
where $\nabla_k$ denotes the covariant derivative with respect to the affine connection $\Gamma^{\,\,k}_{i\,j}$.
The relations (\ref{Ein}), (\ref{con}), (\ref{Car}) and (\ref{sym}) give
\begin{equation}
G^{ik}=\kappa T^{ik}+\frac{1}{2}\kappa^2\biggl(s^{ij}_{\phantom{ij}j}s^{kl}_{\phantom{kl}l}-s^{ij}_{\phantom{ij}l}s^{kl}_{\phantom{kl}j}-s^{ijl}s^k_{\phantom{k}jl}+\frac{1}{2}s^{jli}s_{jl}^{\phantom{jl}k}+\frac{1}{4}g^{ik}(2s^{\phantom{j}l}_{j\phantom{l}m}s^{jm}_{\phantom{jm}l}-2s^{\phantom{j}l}_{j\phantom{l}l}s^{jm}_{\phantom{jm}m}+s^{jlm}s_{jlm})\biggr),
\label{GR}
\end{equation}
where $G_{ik}$ is the Einstein tensor of general relativity.
The second term on the right of (\ref{GR}) is the correction to the curvature of spacetime from the spin \cite{HHKN,HHK}.
If the spin vanishes, (\ref{GR}) reduces to the standard Einstein equations.

\section{Spin fluids}
Quarks and leptons, that compose all stars, are fermions described in relativistic quantum mechanics by the Dirac equation.
Since Dirac fields couple minimally to the torsion tensor, the torsion of spacetime at microscopic scales does not vanish in the presence of fermions \cite{HHKN}.
At macroscopic scales, such particles can be averaged and described as a Weyssenhoff spin fluid \cite{WR,NSH}.
If the spin orientation of particles is random then the macroscopic spacetime average of the spin and of the spin gradients vanish.
On the contrary, the terms that are quadratic in the spin tensor do not vanish after averaging \cite{HHK}.
However, the correction to the curvature from the spin in (\ref{GR}) is significant only at densities of matter much larger than the density of nuclear matter because of the factor $\kappa^2$.

The macroscopic canonical energy-momentum tensor of a spin fluid is given by
\begin{equation}
\sigma_{ij}=c\Pi_i u_j-p(g_{ij}-u_i u_j),
\label{enmo}
\end{equation}
while its macroscopic canonical spin tensor is given by
\begin{equation}
s_{ij}^{\phantom{ij}k}=s_{ij}u^k,\,\,\,s_{ij}u^j=0,
\label{spin}
\end{equation}
where $\Pi_i$ is the four-momentum density of the fluid, $u^i$ its four-velocity, $s_{ij}$ its spin density, and $p$ its pressure \cite{HHK}.
The relations (\ref{sym}), (\ref{GR}), (\ref{enmo}) and (\ref{spin}) give \cite{HHK}
\begin{equation}
G^{ij}=\kappa\Bigl(\epsilon-\frac{1}{4}\kappa s^2\Bigr)u^i u^j-\kappa\Bigl(p-\frac{1}{4}\kappa s^2\Bigr)(g^{ij}-u^i u^j)-\frac{1}{2}\kappa(\delta^l_k+u_k u^l)\nabla_l^{\{\}}(s^{ki}u^j+s^{kj}u^i),
\label{com}
\end{equation}
where $\epsilon=c\Pi_i u^i$ is the rest energy density of the fluid,
\begin{equation}
s^2=\frac{1}{2}s_{ij}s^{ij}>0
\end{equation}
is the square of the spin density, and $\nabla_k^{\{\}}$ denotes the general-relativistic covariant derivative with respect to the Christoffel symbols $\{^{\,\,k}_{i\,j}\}$.
If the spin orientation of particles in a spin fluid is random then the last term on the right of (\ref{com}) vanishes after averaging.
Thus the Einstein-Cartan equations for such a spin fluid are equivalent to the Einstein equations for a perfect fluid with the effective energy density $\epsilon-\kappa s^2/4$ and the effective pressure $p-\kappa s^2/4$ \cite{HHK,Gas,NP,BF}.

\section{Friedman equations with torsion}
A closed, homogeneous and isotropic universe is described by the Friedman-Lema\^{i}tre-Robertson-Walker (FLRW) metric which, in the isotropic spherical coordinates, is given by $ds^2=c^2 dt^2-\frac{a^2(t)}{(1+kr^2/4)^2}(dr^2+r^2 d\theta^2+r^2\mbox{sin}^2\theta d\phi^2)$, where $a(t)$ is the scale factor and $k=1$, and the energy-momentum tensor in the rest frame: $u^0=1,\,u^\alpha=0$ ($\alpha=1,2,3$) \cite{LL}.
The Einstein field equations (\ref{com}) for this metric turn into the Friedman equations \cite{Kop,Tra,dSS1,dSS2,Gas}:
\begin{eqnarray}
& & {\dot{a}}^2+1=\frac{1}{3}\kappa\Bigl(\epsilon-\frac{1}{4}\kappa s^2\Bigr)a^2, \label{Fri1} \\
& & {\dot{a}}^2+2a\ddot{a}+1=-\kappa\Bigl(p-\frac{1}{4}\kappa s^2\Bigr)a^2,
\label{Fri2}
\end{eqnarray}
where the dot denotes the differentiation with respect to $ct$.
These equations yield the conservation law
\begin{equation}
\frac{d}{dt}\bigl((\epsilon-\kappa s^2/4)a^3\bigr)+(p-\kappa s^2/4)\frac{d}{dt}(a^3)=0,
\label{law}
\end{equation}
which can be used instead of the second Friedman equation (\ref{Fri2}).

The average particle number density in a fluid, $n$, is related to the energy density and pressure of the fluid by $dn/n=d\epsilon/(\epsilon+p)$.
If the fluid is described by a barotropic equation of state $p=w\epsilon$ then $n\propto\epsilon^{1/(1+w)}$.
The square of the spin density for a fluid consisting of fermions with no spin polarization is given by
\begin{equation}
s^2=\frac{1}{8}(\hbar cn)^2,
\label{ome1}
\end{equation}
which yields $s^2\propto\epsilon^{2/(1+w)}$ \cite{NP}.
Substituting this relation into (\ref{law}) gives \cite{Gas}
\begin{equation}
\epsilon\propto a^{-3(1+w)},
\label{scaling}
\end{equation}
which has the same form as for $s^2=0$ (in the absence of spin).
Accordingly, the spin-density contribution to the total effective energy density in (\ref{Fri1}) scales like
\begin{equation}
\epsilon_S=-\frac{1}{4}\kappa s^2\propto a^{-6},
\label{ome2}
\end{equation}
regardless of the value of $w$.
Such a scaling is consistent with the conservation of the number of particles, $n\propto a^{-3}$.
Thus $\epsilon_S$ decouples from $\epsilon$ in time evolution of the Universe.
The spin fluid can be regarded as a mixture of a fluid in the standard Friedman equations and an exotic fluid for which $p=\epsilon<0$.
This picture is purely formal, however, because such an exotic fluid cannot exist alone and its relation $p=p(\epsilon)$ does not represent a physical equation of state.

In the very early Universe which we consider, Dirac particles composing the spin fluid had energies much greater than their rest energies.
Thus they are described by an ultrarelativistic barotropic equation of state, $p=\epsilon/3$ ($w=1/3$), as for radiation.
Since the relic background photons and neutrinos are the most abundant particles in the Universe \cite{Ric}, the energy density of a spin fluid in the very early Universe was $\epsilon\approx\epsilon_R\approx\epsilon_\gamma+\epsilon_\nu$, where $R$ denotes radiation, $\gamma$ photons, and $\nu$ neutrinos.
The scaling relation (\ref{scaling}) gives $\epsilon_R\propto a^{-4}$.
The total effective energy density is then given by
\begin{equation}
\epsilon+\epsilon_S=\epsilon_{R0}\hat{a}^{-4}+\epsilon_{S0}\hat{a}^{-6},
\label{tot}
\end{equation}
where $\hat{a}=a/a_0$ is the normalized scale factor and the subscripts $0$ denote quantities measured at the present time (when $\hat{a}=1$).
We neglect the contributions to (\ref{tot}) from matter (dark plus baryonic), $\epsilon_{M0}\hat{a}^{-3}$, and from the cosmological constant, $\epsilon_\Lambda=\Lambda/\kappa$, because in the early Universe ($\hat{a}\ll 1$) they become much smaller than $\epsilon_{R0}\hat{a}^{-4}$ \cite{Ric}.

The first Friedman equation (\ref{Fri1}) can be written as
\begin{equation}
H^2+\frac{c^2}{a^2}=\frac{1}{3}\kappa(\epsilon+\epsilon_S)c^2,
\label{Fri}
\end{equation}
where $H=c\dot{a}/a$ is the Hubble parameter.
The present-day total density parameter, $\Omega=(\epsilon_0+\epsilon_{S0})/\epsilon_c$, where $\epsilon_c=3H^2_0/(\kappa c^2)$ is the present-day critical energy density, gives $a_0 H_0\sqrt{\Omega-1}=c$, as in general-relativistic cosmology \cite{Ric}.
The total density parameter at any instant,
\begin{equation}
\Omega(\hat{a})=\frac{\kappa c^2}{3H^2}(\epsilon+\epsilon_S),
\end{equation}
satisfies
\begin{equation}
a|H|\sqrt{\Omega(\hat{a})-1}=c,
\label{ide}
\end{equation}
Using the present-day density parameters,
\begin{equation}
\Omega_R=\epsilon_{R0}/\epsilon_c,\,\,\,\Omega_S=\epsilon_{S0}/\epsilon_c,
\end{equation}
in (\ref{tot}) brings (\ref{Fri}) to $|H|=H_0\bigl(\Omega_R\hat{a}^{-4}+\Omega_S\hat{a}^{-6}-(\Omega-1)\hat{a}^{-2}\bigr)^{1/2}$.
The last term in the above equation is much smaller than the terms with $\Omega_R$ and $\Omega_S$ if $\hat{a}\ll 1$, thus it can be neglected, leading to
\begin{equation}
|H|=H_0\Bigl(\Omega_R\hat{a}^{-4}+\Omega_S\hat{a}^{-6}\Bigr)^{\frac{1}{2}},
\label{Hub}
\end{equation}
which shows how the Hubble parameter depends on $\hat{a}$ in the early Universe.
The relations (\ref{ide}) and (\ref{Hub}) give the total density parameter as a function of $\hat{a}$:
\begin{equation}
\Omega(\hat{a})=1+\frac{(\Omega-1)\hat{a}^4}{\Omega_R\hat{a}^2+\Omega_S}.
\label{omega}
\end{equation}
Since the energy-density contribution from torsion $\epsilon_S$ is negative, so is the torsion density parameter $\Omega_S$.
This contribution generates gravitational repulsion which is significant at very small $\hat{a}$.

\section{Density parameters}
Seven-year Wilkinson Microwave Anisotropy Probe (WMAP) observations show that our Universe may be indeed closed, with $\Omega=1.002$ \cite{WMAP}.
The WMAP data give also $H^{-1}_0=4.4\times10^{17}\,\mbox{s}$ and $\Omega_R=8.8\times10^{-5}$.
Thus $a_0=2.9\times10^{27}\,\mbox{m}$.
To estimate $\Omega_S$, we use the relic background neutrinos which are the most abundant fermions in the Universe, with $n=5.6\times10^7\,\mbox{m}^{-3}$ for each type (out of 6) \cite{Ric}.
Equations (\ref{ome1}) and (\ref{ome2}) then give
\begin{equation}
\Omega_S=-8.6\times10^{-70}.
\label{omegas}
\end{equation}
While in general relativity the torsion density parameter $\Omega_S$ vanishes, the ECKS theory of gravity gives $\Omega_S$ a nonzero, though extremely small, negative value.

\section{Flatness problem}
Gravitational repulsion induced by torsion, which becomes significant at extremely high densities, prevents the cosmological singularity.
Equation (\ref{Hub}) shows that the expansion of the Universe started when $H=0$, at which $\hat{a}=\hat{a}_m$, where
\begin{equation}
\hat{a}_m=\sqrt{-\frac{\Omega_S}{\Omega_R}}=3.1\times10^{-33},
\label{mini}
\end{equation}
corresponding to the minimum but finite scale factor (radius of a closed universe) $a_m=9\times10^{-6}\,\mbox{m}$.
Before reaching its minimum size, the Universe was contracting with $H<0$.
If we choose $t=0$ at $\hat{a}=\hat{a}_m$ then integrating (\ref{Hub}) for $t>0$ gives
\begin{equation}
-\frac{\Omega^{3/2}_R H_0}{\Omega_S}t=f(x)=\frac{x}{2}\sqrt{x^2-1}+\frac{1}{2}\mbox{ln}|x+\sqrt{x^2-1}|,
\end{equation}
where $x=\hat{a}/\hat{a}_m$.
When $x\gg1$, $f(x)\approx x^2/2$, yielding the usual evolution of the radiation-dominated Universe, $a\sim t^{1/2}$.

In general relativity, $\Omega_S=0$, from which it follows that $\Omega(\hat{a})$ in (\ref{omega}) tends to 1 as $\hat{a}\rightarrow 0$ according to $\Omega(\hat{a})-1=(\Omega-1)\hat{a}^2/\Omega_R$, which introduces the flatness problem in Big-Bang cosmology.
$\Omega(\hat{a})$ at the GUT epoch must have been tuned to 1 to a precision of more than 52 decimal places in order for $\Omega$ to be near 1 today.
This problem can be solved by cosmic inflation, according to which the Universe in the very early stages of its evolution exponentially expanded (which involved false vacuum or scalar fields) by a factor of at least $10^{26}$, making $\Omega(\hat{a})$ sufficiently close to 1 at the GUT epoch \cite{inf1,inf2}.

In the ECKS gravity, where $\Omega_S<0$, $\Omega(\hat{a})$ is infinite at $\hat{a}=\hat{a}_m$.
The function (\ref{omega}) has a local minimum at $\hat{a}=\sqrt{2}\hat{a}_m$, where it is equal to
\begin{equation}
\Omega(\sqrt{2}\hat{a}_m)=1-\frac{4\Omega_S(\Omega-1)}{\Omega^2_R}=1+8.9\times10^{-64}.
\label{min}
\end{equation}
As the Universe expands from $\hat{a}_m$ to $\sqrt{2}\hat{a}_m$, $\Omega(\hat{a})$ rapidly decreases from infinity to the value (\ref{min}) which {\em appears} to be tuned to 1 to a precision of about 63 decimal places.
This stage takes
\begin{equation}
t=-\frac{\Omega_S}{\Omega^{3/2}_R H_0}f(\sqrt{2})=5.3\times10^{-46}\,\mbox{s}.
\label{time}
\end{equation}
During this time, the Universe expands only by a factor of $\sqrt{2}$ which is much less than $10^{26}$ in the inflationary scenario.
Thus the apparent fine tuning of $\Omega(\hat{a})$ in the very early Universe is naturally caused by the {\em extremely small and negative torsion density parameter} (\ref{omegas}) originating from the torsion of spacetime in the ECKS gravity, without needing the inflationary dynamics.
As the Universe expands further, $\Omega_R\hat{a}^2$ becomes much greater than $|\Omega_S|$ and $\Omega(\hat{a})-1$ increases according to $\Omega(\hat{a})-1=(\Omega-1)\hat{a}^2/\Omega_R$, until the Universe becomes dominated by matter.
In this epoch, $w<1/3$ and the contributions from dark and baryonic matter and from the cosmological constant must be included in (\ref{tot}).

\section{Horizon problem}
The relations (\ref{ide}) and (\ref{min}) give
\begin{equation}
\dot{a}=\frac{1}{\sqrt{\Omega(\hat{a})-1}}.
\end{equation}
The velocity of the point that is antipodal to the coordinate origin, $v_a=\pi c\dot{a}$ \cite{LL,Ric}, has a local maximum at $\hat{a}=\sqrt{2}\hat{a}_m$, where it is equal to
\begin{equation}
v_a=\frac{\pi\Omega_R}{2\sqrt{-\Omega_S(\Omega-1)}}c=1.1\times10^{32}\,c.
\label{max}
\end{equation}
As the closed Universe expands from $\hat{a}_m$ to $\sqrt{2}\hat{a}_m$, $v_a$ rapidly increases from zero to the enormous value (\ref{max}).
During this time, the Universe is accelerating: $\ddot{a}>0$.
As the Universe expands further, $v_a$ decreases according to $v_a=\pi c\sqrt{\Omega_R}\hat{a}^{-1}/\sqrt{\Omega-1}$, until the Universe becomes dominated by matter, when the formula for the decrease of $v_a$ depends also on the contributions from dark and baryonic matter (and later also from the cosmological constant).
During this time, the Universe is decelerating: $\ddot{a}<0$, until the cosmological constant becomes dominant and the Universe is accelerating again.

If the closed Universe was causally connected at some instant $t<0$, then it remains causally connected during its contraction until $t=0$ and also during the subsequent expansion until $v_a$ reaches $c$.
After that moment, the point at the origin cannot communicate with points in space receding with velocities greater than $c$.
That is, the Hubble radius $d_H=c/H$ becomes smaller than the physical distance to the antipodal point $d_a=\pi a$.
The Universe contains $N\approx(v_a/c)^3=(d_a/d_H)^3$ causally disconnected volumes.
At $t$ given by (\ref{time}), $d_a$ is 32 orders of magnitude greater than $d_H$ and $N\approx10^{96}$.
Again, it is the {\em extremely small and negative torsion density parameter} (\ref{omegas}) that naturally causes how such a large number of causally disconnected volumes arises from a single causally connected region of spacetime, without needing the inflationary dynamics.
As the Universe expands further, $|\Omega_S|$ becomes negligible, the Universe smoothly enters the radiation-domination epoch, and $N$ decreases according to standard cosmology.

\section{Discussion}
In this work, we considered the ECKS theory of gravity which is the closest theory with torsion to general relativity.
We used the spin density of matter as the source of torsion, which has a natural physical interpretation in the context of the Poincar\'{e} group and does not introduce additional fields or coupling constants \cite{HHKN}.
We used the spin-fluid form of the spin density, which can be derived from the conservation law for the spin density using the method of multipole expansion \cite{Niko0,NSH}.
Although the spin is usually considered as the source of torsion, there exist several other possibilities in which torsion can emerge.
Different forms of the torsion tensor can arise from such sources, giving different modifications to the energy-momentum tensor \cite{Cap}.
These torsion fields could be used to derive the same effects on the geometry of the Universe as those caused by the torsion from a spin fluid.
Whether torsion couples to spin and whether other kinds of torsion are physical should be tested experimentally.
Many gravitational Lagrangians with torsion (including the ECKS one) imply that the torsion is related to its source through an algebraic equation, so that the torsion does not propagate and vanishes in vacuum, which introduces limitations on its detection \cite{Sha,Ham}.
If torsion can exist in vacuum but it couples only to intrinsic spin, its effects in the Solar System would be still negligibly small \cite{HHKN}.
In this case, typical experimental limits on torsion come from searches for dynamical properties of particles such as quantum effects from the coupling of torsion to Dirac spinors, spin-spin interaction due to torsion or anomalies in the Standard Model with torsion \cite{Sha,Ham}.
However, if torsion couples to rotational angular momentum, such a coupling would affect the physical phenomena at larger scales, including the precession of gyroscopes without spin polarization such as a gyroscope in the Gravity Probe B experiment \cite{Mao}.

The physical picture for a spin fluid considered in this Letter is the same as that in Gasperini's model of spin-dominated inflation \cite{Gas} for $w=1/3$.
It has been shown in \cite{Gas} that a spin fluid with $w=-1/3+\delta$, where $\delta>0$ is a number extremely close to zero, can characterize a physically viable inflationary scenario.
The origin of such a fine tuning would be difficult to explain, though.
We consider the physically realistic case $w=1/3$ and analyze $\Omega(\hat{a})$ to show that an extremely small and negative torsion density parameter $\Omega_S\approx -10^{-69}$, arising from a very weak and repulsive spin-spin interaction predicted by the ECKS theory of gravity, {\em suffices to solve the flatness and horizon problems without any fine tuning}.
The ECKS cosmology is thus a compelling alternative not only to Big-Bang cosmology (it avoids the initial singularity), but also to the standard inflationary scenario because it does not require false vacuum, scalar fields or other speculative mechanisms.
Moreover, the ECKS gravity does not contain free parameters ($G$ and $c$ can be always set to 1 by changing the units), thus it is advantageous over inflationary models that do introduce new parameters \cite{inf2}.

According to (\ref{Hub}), the contraction of the Universe before $t=0$ looks like the time reversal of the following expansion.
However, the idea of a universe contracting from infinity in the past does not explain what caused such a contraction, just like Big-Bang cosmology cannot explain what happened before the Big Bang.
Fortunately, two mechanisms can cause the dynamics asymmetry between the contraction and expansion.
First, when the Universe has the minimum radius (\ref{mini}), the radiation energy density is $\epsilon_R=1.1\times10^{116}\,\mbox{J}\,\mbox{m}^{-3}$, which is greater than the Planck energy density by a few orders of magnitude.
Thus it is also significantly greater than the density threshold for particle production \cite{HHKN,part1,part2,part3}.
Such pair production would increase $\Omega_S$.
If the contracting Universe was anisotropic in the past, the particle production in the presence of extremely large tidal forces would also increase the energy density to the values sufficient for isotropization of the subsequent expansion \cite{part1,part2}.
Second, the electroweak interactions between fermions in the early Universe could cause their spins to align, making the last term on the right of (\ref{com}) nonzero.
This term would introduce in the Friedman equations a time asymmetry with respect to the transformation $t\rightarrow-t,\,H\rightarrow-H$.
The presence of the covariant derivatives of the four-velocity in this term could contribute to the production of mass in the Universe \cite{Lord}.
Also, the spin tensor in this term acts like viscosity, which would increase the entropy of the Universe.

We propose the following scenario.
A massive star, that is causally connected, collapses gravitationally to a black hole and an event horizon forms.
Inside the horizon, spacetime is nonstationary and matter contracts to an extremely dense, but because of torsion, finite-density state.
In the frame locally moving with matter, this contraction looks like the contraction of a closed universe \cite{LL,Tol}.
Such a universe is initially causally connected and anisotropic.
Extremely large tidal forces cause an intense pair production which generates the observed amount of mass and increases the energy density, resulting in isotropization of this universe \cite{part1,part2,part3}.
Additional terms in the Lagrangian density containing torsion could also generate massive vectors \cite{Niko3}.
The spin density increases, magnifying the repulsive spin-fluid forces due to the negative $\epsilon_S$.
The particle production does not change the total (matter plus gravitational field) energy of the resulting FLRW universe, which is zero if we neglect the cosmological constant \cite{CI}.
After reaching its minimum size, the homogeneous and isotropic universe starts expanding.
Such an expansion is not visible for observers outside the black hole, for whom the horizon's formation and all subsequent processes occur after infinite time \cite{LL}.
The new universe is thus a separate spacetime branch with its own timeline; it can last infinitely long and grow infinitely large if dark energy is present.

As the universe in a black hole expands to infinity, the boundary of the black hole becomes an Einstein-Rosen bridge connecting this universe with the outer universe \cite{ER}.
We recently suggested that all astrophysical black holes may be Einstein-Rosen bridges (wormholes), each with a new universe inside that formed simultaneously with the black hole \cite{Niko1}.
Accordingly, our own Universe may be the interior of a black hole existing in another universe, and the time asymmetry of motion at the boundary of this black hole may cause the perceived arrow of cosmic time.
This scenario is possible because the torsion of spacetime, which is produced by the intrinsic spin of fermions in a spin fluid, prevents the formation of singularities \cite{HHK}.
Thus the gravitational collapse of a star composed of quarks and leptons to a black hole does not create a singularity \cite{Niko2}, allowing matter inside the event horizon to undergo a bounce at the (nonsingular) state of maximum density and then reexpand.
A similar scenario, where a new universe emerges inside each collapsing black hole, has been proposed earlier by Smolin \cite{Smo1,Smo2}, however, without providing a physical mechanism for such a bounce and with invoking inflation.
It has also been argued in \cite{EB} that the horizon, flatness, and structure formation problems can be solved without inflation if the (flat, with $k=0$) Universe is born from the interior of a black hole, however, without demonstrating how to avoid the formation of a central singularity in a collapsing black hole.
Torsion thus provides a viable theoretical explanation of how black-hole interiors can generate new universes, avoiding unphysical singularities and replacing speculative mechanisms of inflation.
Cosmology in a black hole has also been proposed in \cite{BGG,ST}.

Since most stars rotate, most astrophysical black holes are rotating black holes.
A universe born from a rotating black hole should inherit its preferred direction, related to the axis of rotation.
Such a preferred direction should introduce small corrections to the FLRW metric, containing the Kerr radius $a=M/(mc)$, where $M$ is the angular momentum of a rotating black hole and $m$ is its mass \cite{LL,Kerr}.
These corrections could then couple to other fields, allowing to verify whether our Universe was born in a black hole.
GRS 1915+105, which is the heaviest and fastest spinning, known stellar black hole in the Milky Way Galaxy, has $a<26\,\mbox{km}$ \cite{puls}.
Lighter or slower spinning black holes have smaller values of $a$.
To compare, the preferred-frame parameter $2.4\times10^{-19}\,\mbox{GeV}$ in a model for neutrino oscillations using Lorentz violation \cite{Tep} corresponds to the length of $820\,\mbox{m}$.

The proposed scenario for the origin of our Universe may explain the arrow of time \cite{Smo1}.
Although the laws of the ECKS theory of gravity are time-symmetric, the boundary conditions of the Universe are not, because the motion of matter through the event horizon of a black hole is unidirectional and thus it can define the arrow of time.
The arrow of cosmic time of a universe inside a black hole would then be fixed by the time-asymmetric collapse of matter through the event horizon, before the subsequent expansion.
Such an arrow of time would also be entropic: although black holes are states of maximum entropy in the frame of outside observers, new universes expanding inside black holes would allow entropy to increase further.

\section*{Acknowledgements}
I am grateful to Chris Cox for his support and help in writing this Letter.
I dedicate this work to my Parents.

\end{document}